\documentclass[twocolumn,showpacs,preprintnumbers]{revtex4}%
\usepackage{amsfonts}
\usepackage{amsmath}
\usepackage{graphicx}
\usepackage{dcolumn}
\usepackage{bm}
\usepackage{amssymb}
\usepackage{slashed}
\usepackage{acronym}%
\begin{document}
\title{Restoration of chiral symmetry in the large-$N_{c}$ limit}
\author{Achim Heinz$^{\text{(a)}}$, Francesco Giacosa$^{\text{(a)}}$, and Dirk H.\
Rischke$^{\text{(a,b)}}$}
\affiliation{$^{\text{(a)}}$Institute for Theoretical Physics, Goethe University,
Max-von-Laue-Str.\ 1, D--60438 Frankfurt am Main, Germany }
\affiliation{$^{\text{(b)}}$Frankfurt Institute for Advanced Studies, Ruth-Moufang-Str.\ 1,
D--60438 Frankfurt am Main, Germany }

\begin{abstract}
We study the large-$N_{c}$ behavior of the critical temperature $T_{c}$ for
chiral symmetry restoration in the framework of the
Nambu--Jona-Lasinio (NJL) model and the linear $\sigma$-model. While in
the NJL case $T_{c}$ scales as $N_{c}^{0}$ and is, as expected, of the same
order as $\Lambda_{QCD}$ (just as the deconfinement phase transition), in the
$\sigma$-model the scaling behavior reads $T_{c}\propto N_{c}^{1/2}$. We
investigate the origin of the different scaling behavior 
and present two improvements of the $\sigma$-model: (i) 
a simple, phenomenologically motivated
temperature dependence of the parameters and (ii) the coupling to the
Polyakov loop. Both approaches lead to the scaling $T_{c}\propto
N_{c}^{0}$.

\end{abstract}

\maketitle



Although quantum chromodynamics (QCD) is a theory for quarks and
gluons, in the vacuum they are confined inside hadrons. It is,
however, expected that at sufficiently high temperature and/or density a phase
transition to a deconfined gas of interacting quarks and gluons takes place
\cite{Cabibbo:1975ig,Rischke:2003mt}. Moreover, it is also expected that this
deconfinement phase transition is related to the so-called chiral
phase transition: chiral symmetry is broken in the vacuum and restored in a
hot and/or dense medium, see Ref.\ \cite{Casher:1979vw}
and the lattice simulations of Refs.\ \cite{Cheng:2006qk,Aoki:2006br}.

The precise connection between the deconfinement and the chiral phase
transition is not yet clear. This is also due to the fact that both
transitions can only be precisely defined in limiting situations which are not
realized in nature. Namely, in the context of pure Yang-Mills theory (QCD with
infinitely heavy quarks) the Polyakov loop is the order parameter for the
deconfinement phase transition \cite{Polyakov:1978vu}: the expectation value
of the Polyakov loop vanishes for low $T$ and $\mu$ (confined matter) and
approaches unity in the deconfined phase. On the other hand, in the limit of zero
quark masses the QCD Lagrangian is invariant under chiral
transformations. The chiral condensate $\left\langle \bar{q}q\right\rangle
$ is the order parameter for the chiral phase transition: it is nonzero in the
vacuum, decreases for increasing $T$ and $\mu$, and vanishes in the
chirally restored phase. Nature is somewhere in between: the quark masses are
neither infinite nor zero. The chiral condensate and the Polyakov
loop are therefore only approximate order parameters.

Since QCD cannot be directly solved, various methods are used to perform
explicit calculations. Besides the already mentioned lattice simulations,
effective models containing quark degrees of freedom only, such as the 
NJL model
\cite{Nambu:1961tp,Klevansky:1992qe,Vogl:1991qt,Hatsuda:1994pi}, and purely
hadronic models, such as the linear $\sigma$-model
\cite{Gell-Mann:1960ls,Parganlija:2010fz}, have been used to study the
thermodynamics of QCD. Both approaches cannot describe the deconfinement phase
transition: the degrees of freedom of NJL models are deconfined quarks
at all temperatures and densities, while linear $\sigma$-models 
feature hadronic degrees of freedom in which
quarks are always confined. However, in both approaches 
the critical temperature for the chiral phase
transition is in agreement with recent lattice studies.

In order to amend these problems, generalizations of the NJL-model have been
developed recently, in which the Polyakov loop has been coupled to the quarks
\cite{Ratti:2005jh,Fukushima:2008wg} (see also Ref.\ \cite{Schaefer:2007pw}, in
which --besides quarks-- also mesons are present). Although confinement is
not realized in a strict sense because mesons can still decay into
quark-antiquark pairs \cite{Hansen:2006ee}, an effective description of
confinement is achieved through the behavior of the Polyakov loop at nonzero $T$
and $\mu$. Thus, both deconfinement and chiral phase transition can be
studied in a unified framework.

Another widely used approach to study QCD both in the vacuum and in the medium
is the so-called large-$N_{c}$ limit \cite{tHooft:1973jz,Witten:1979kh} in
which the number of colors $N_{c}$ is not fixed to $3$, but is sent to infinity.
When enlarging the gauge symmetry of QCD from $SU(3)$ to $SU(N_{c})$
with $N_c\gg3$ one
obtains a theory which still contains mesons and baryons, but is --although
not solvable-- substantially simpler \cite{Coleman:1980mx}. Relevant
quantities can be expressed as a series in $N_{c}^{-n}$, so that it is possible
to separate large-$N_{c}$ dominant and large-$N_{c}$ suppressed terms. 
This is the only known approach to understand some phenomenologically
well-established properties of QCD such as the Zweig rule. In order for the
large-$N_{c}$ limit to be consistent, the following scaling of the QCD
coupling $g_{QCD}$ must be implemented:
\begin{equation}
N_{c}\rightarrow\infty~~,~~g_{QCD}^{2}~N_{c}\rightarrow\text{finite .}%
\end{equation}
In this way quark-antiquark meson masses scale with $N_{c}^{0}$ and the
amplitude for a $k$-leg quark-antiquark interaction vertex scales as
$N_{c}^{-(k-2)/2}$ and thus goes to zero for $N_{c}\rightarrow\infty.$ In
particular, decay amplitudes are suppressed as $1/\sqrt{N_{c}}$ and therefore
quark-antiquark mesons are stable and non-interacting in the large-$N_{c}$
limit. Thus, at nonzero $T$ a non-interacting gas of mesons is realized for
$N_{c}$ $\gg3$.

It is expected that the deconfinement and the chiral critical temperatures are
independent of the number of colors, see Refs.\
\cite{Thorn:1980iv,McLerran:2007qj} and refs.\ therein. In fact, they should be
proportional to the only existing QCD scale, $\Lambda_{QCD}\propto N_{c}^{0}$.
In this work we aim to study this issue in detail: namely, we investigate the
large-$N_{c}$ behavior of the chiral phase transition in the previously
mentioned NJL and linear $\sigma$-model. Quite remarkably, we find that
these models behave very differently: while in the NJL model the chiral phase
transition $T_{c}$ scales as $N_{c}^{0},$ as expected, in the linear $\sigma
$-model $T_{c}$ scales as $\sqrt{N_{c}}$ and is thus not consistent
with the above expectations and with the NJL result. Note that these results,
although proven within the simplest possible versions for both the NJL and the
linear $\sigma$-model ($N_{f}=2,$ no vector mesons, etc.), are based only on
general large-$N_{c}$ properties and therefore hold also when considering more
complicated and realistic generalizations. 

The reason for the (inconsistent) behavior of the linear $\sigma$-models at
large-$N_{c}$ is investigated and two improvements are discussed:
(i) a phenomenologically based modification of the model, in which (at least)
one coupling constant becomes explicitly temperature-dependent; (ii) the
coupling of the $\sigma$-model to the Polyakov-loop degree of freedom. In both
ways the correct limit $T_{c}\propto N_{c}^{0}$ is recovered. Thus, it is
still possible to use chiral hadronic models for studying the chiral phase
transition, although the present study shows that some modifications are
needed in order to be consistent with the large-$N_{c}$ limit.


The NJL-model is a quark-based chiral model
\cite{Nambu:1961tp,Klevansky:1992qe,Vogl:1991qt,Hatsuda:1994pi}, which has
been widely used to study the chiral phase transition in the medium. It is
based on a chirally symmetric four-quark point-like interaction.
The Lagrangian in the case $N_{f}=2$ reads as function of $N_{c}$:
\begin{equation}
\mathcal{L}_{NJL}(N_{c})=\bar{\psi}(\imath\gamma^{\mu}\partial_{\mu}%
-m_{q})\psi+\frac{3G}{N_{c}}\left[  (\bar{\psi}\psi)^{2}+(\bar{\psi}%
\imath\gamma_{5}\psi)\right]  ^{2}\text{ ,}%
\end{equation}
where $\psi^{t}=(u,d)$ is the quark spinor, $m_{q}$ is the bare quark mass and
$G$ is the coupling constant with dimension energy$^{-2}$, whose $N_c$-scaling
$G\rightarrow3G/N_{c}$ (following from the relation $G\propto g_{QCD}^{2}$)
has been made explicit. The quark develops a constituent mass $m^{\ast}$
which is proportional to the chiral condensate $\left\langle \bar
{q}q\right\rangle $: $m^{\ast}=-2G\left\langle \bar{q}q\right\rangle
.$ 
In mean-field approximation
the effective mass $m^{\ast}$ as a function of $T$ and $N_{c}$ reads
\cite{Klevansky:1992qe}
\begin{equation} \label{mast}
1=\frac{m_{q}}{m^{\ast}}+\frac{3G}{N_{c}}\left(  2N_{c}+\frac{1}{2}\right)
\int_{0}^{\Lambda}\frac{dk~k^{2}}{\pi^{2}}\frac{2 \tanh\left(  \frac
{\sqrt{k^{2}+m^{\ast2}}}{2T}\right)  }{\sqrt{k^{2}+m^{\ast2}}}\text{ ,}%
\end{equation}
where a cutoff $\Lambda$ has been introduced in order to regularize the
loop integral. Note that the number $N_c$ of quarks running in the loop cancels
with the factor $1/N_c$ from the coupling constant, such that 
the dominant term in Eq.\ (\ref{mast}) is independent of $N_c$.
In the chiral limit $m_{q}\rightarrow0$
the critical temperature for chiral symmetry restoration $T_{c}$ 
is obtained as the temperature at which the effective
mass $m^{\ast}$, and therefore also the chiral condensate $\left\langle
\bar{q}q\right\rangle $, vanish. To leading order in $N_{c}$ it reads
\begin{equation}
T_{c}(N_{c})\simeq\Lambda\sqrt{\frac{3}{\pi^{2}}}\sqrt{1-\frac{ \pi^{2}%
}{6\Lambda^{2}G}}\text{ }\propto N_{c}^{0}\text{ .}\label{tcnjl}%
\end{equation}
When $m_{q}>0,$ a
crossover is realized and the corresponding (pseudo-)critical temperature, defined as
the point at which the first derivative $\left\vert dm^{\ast}/dT\right\vert $
is maximal, is also $N_{c}$-independent. This result, based on general
scaling arguments, does not change when including the $s$-quark and
(axial-)vector degrees of freedom. Moreover, it is also unaffected by the 't
Hooft terms describing the $U_{A}(1)$ anomaly which is suppressed in the
large-$N_{c}$ limit. We thus conclude that in all versions of the NJL
model the critical temperature for chiral symmetry restoration (second order or
crossover) is independent of the number of colors. It has therefore, as
expected, the same scaling as the deconfinement phase transition.


The linear $\sigma$-model is a purely hadronic theory constructed from the
requirements of chiral symmetry and its spontaneous breaking
\cite{Gell-Mann:1960ls,Giacosa:2006tf,Parganlija:2010fz}, out of which the
pions emerge as Goldstone bosons in the chiral limit.

In order to study the large-$N_{c}$ behavior of the chiral phase transition we
consider, as in the NJL model, the case $N_{f}=2$ in the chiral limit, in
which the Lagrangian as function of $N_{c}$ reads:
\begin{equation}
\mathcal{L}_{\sigma}(N_{c})=\frac{1}{2}(\partial_{\mu}\Phi)^{2}+\frac{1}{2}%
\mu^{2}\Phi^{2}-\frac{\lambda}{4}\frac{3}{N_{c}}\Phi^{4}\text{ ,}\label{ls}%
\end{equation}
where $\Phi^{t}=(\sigma,\vec{\pi})$ describes the scalar field $\sigma$ and
the pseudoscalar pion triplet $\vec{\pi}$ and where the standard scaling
behaviors $\lambda\rightarrow\frac{3}{N_{c}}\lambda$ (suppression of the
interaction) and $\mu^{2}\rightarrow\mu^{2}$ (constancy of meson masses) have
been implemented. For $\mu^{2}>0$ a nonzero chiral condensate $\varphi_{0}%
=\varphi(T=0)=\mu\sqrt{N_{c}/3\lambda}=\sqrt{N_{c}/3}f_{\pi}$ emerges
($f_{\pi}$ is the pion decay constant). The tree-level masses for the sigma and
the pions are $m_{\sigma}^{2}=3\lambda f_{\pi}^{2}-\mu^{2}$ , $m_{\pi}^{2}=0$. 

Many investigations of the linear $\sigma$-model at nonzero $T$ have been performed
in the past, see e.g.\ Refs.\
\cite{Bochkarev:1995gi,Lenaghan:1999si,Petropoulos:2004bt} and refs.\ therein.
Using the Cornwall-Jackiw-Tomboulis (CJT) formalism
\cite{Cornwall:1974vz} in double-bubble approximation and
excluding the trivial solution,
we obtain the gap equation \cite{Lenaghan:1999si} 
\begin{equation} \label{gap}
0=  \varphi(T)^{2}-\frac{N_c}{3 \lambda}\, \mu^2 +
3  \int \left( G_{\sigma}+ G_{\pi}\right) \;.
\end{equation}
The tadpole integrals over the full propagators $G_{\sigma}$ and
$G_{\pi}$ of $\sigma$ and $\pi$ meson read (in the so-called trivial
renormalization scheme where vacuum fluctuations are neglected)
$$
\int G_{i}=\int_{0}^{\infty}\frac{dk~k^{2}}{2\pi^{2}\sqrt{k^{2}+M_{i}^{2}}}
\left[  \exp\left(  \frac{\sqrt{k^{2}+M_{i}^{2}}}%
{T}\right)  -1\right]^{-1}\!\!\!,
$$ 
where $M_i$ is the effective $T$-dependent mass of either $\sigma$
meson or pion. At the critical temperature $T_{c}$
the condensate $\varphi(T)$ and the masses of $\sigma$ and pion vanish.
The gap equation (\ref{gap}) leads to the expression
\begin{equation}
T_{c}(N_{c})=\sqrt{2}f_{\pi}\sqrt{\frac{N_{c}}{3}}\propto N_{c}^{1/2}\text{
.}\label{tcs}%
\end{equation}
For $N_{c}=3$ the critical temperature is, as known, $T_{c}=\sqrt{2}f_{\pi}$
\cite{Bochkarev:1995gi}, but for an infinite number of colors the phase
transition does not take place: for $N_{c}$ $\gg3$ a gas of non-interacting
mesons is realized and the meson loops, which are responsible for chiral
restoration, are suppressed. Note that the scaling behavior of Eq.\
(\ref{tcs}) is not a prerogative of the simple Lagrangian (\ref{ls}), but
holds also in more general hadronic models as those of Ref.\ \cite{Parganlija:2010fz}.

However, Eq.\ (\ref{tcs}) contradicts Eq.\ (\ref{tcnjl}). This mismatch, already
noticed in Ref.\ \cite{Megias:2004hj}, is puzzling because both approaches
contain the same symmetries. Moreover, the linear $\sigma$-model can be obtained as
the hadronized version of the NJL model. However, the hadronization procedure
should be performed for each temperature $T$ and, as a consequence, the
coupling constants in the linear $\sigma$-model should be functions of $T.$ In
particular, the chiral condensate $\varphi(T)$ of the $\sigma$-model should
not be larger than in the corresponding NJL model.
In the following, we discuss two improvements of the linear
$\sigma$-model which repair the mismatch in the $N_c$-scaling
of $T_c$.



We first present a simple and phenomenologically motivated modification of the
linear $\sigma$-model which leads to the correct large-$N_{c}$
results. This
consists of replacing the parameter $\mu^{2}$ with a $T$-dependent function,
\begin{equation}
\mu^{2}\longrightarrow\mu(T)^{2}=\mu^{2}\left(  1-\frac{T^{2}}{T_{0}^{2}}\right)
\text{ ,}\label{musc}%
\end{equation}
where the parameter $T_{0}\simeq\Lambda_{QCD}\propto N_{c}^{0}$ introduces a
new temperature scale. We have implemented the quadratic $T$
dependence suggested in Ref.\ \cite{Gasser:1986vb}. 
Inserting Eq.\ (\ref{musc}) into Eq.\ (\ref{gap}) and following the same steps that
led to Eq.\ (\ref{tcs}), the critical temperature now reads
\begin{equation} \label{tcsmuT}
T_{c}(N_{c})=T_{0}\left(  1+\frac{1}{2}\frac{T_{0}^{2}}{f_{\pi}^{2}}\frac
{3}{N_{c}}\right)  ^{-1/2}\text{ ,}%
\end{equation}
and is independent of $N_{c}$ in the limit $N_{c}\rightarrow\infty$:
$\lim_{N_{c}\rightarrow\infty}T_{c}(N_{c})=T_{0}$. It is also clear how the
meson tadpoles affect the chiral phase transition. Without their contribution, the
transition temperature is simply $T_c = T_0$, like in the
large-$N_c$ limit, and independent of $N_c$. 
The meson tadpoles are responsible for the term $\propto 3/(N_c
f_\pi^2)$ in Eq.\ (\ref{tcsmuT}) and thus lead to a reduction of
$T_c$, $T_c < T_0$, for any finite value of $N_c$.
In the case $N_{c}=3$, using the numerical value
$f_{\pi}=92.4~\text{MeV}$ and setting the temperature scale $T_{0}%
=\Lambda_{QCD}\simeq225~\text{MeV}$, the critical temperature $T_{c}$ is
lowered to $T_{c}\simeq113~\text{MeV}$. Interestingly, in the framework of
sigma models with (axial-)vector mesons, one has to make the replacement
$f_{\pi}\rightarrow Zf_{\pi}$ with $Z\simeq 1.67$ \cite{Parganlija:2010fz}. 
This leads to a critical temperature $T_{c}\simeq157~$MeV, which is remarkably
close to lattice results \cite{Cheng:2006qk,Aoki:2006br}. With the help of
the described modification the linear $\sigma$-model respects the
large-$N_{c}$ limit and is compatible with the NJL model.

Note that we could have also introduced a $T$-dependent coupling
constant $\lambda(T)$ instead of a $T$-dependent mass parameter
$\mu(T)^2$. As long as $\lambda(T)$ does not depend on $N_c$, our conclusions remain
unchanged. Note also that the tadpoles in Eq.\ (\ref{gap}) are
natural candidates to induce the quadratic $T$-dependence of $\mu(T)^2$
in Eq.\ (\ref{musc}). However, they are proportional to $1/N_c$, while
in our case we have to require that the loop contributions 
reponsible for the $T$-dependence in Eq.\ (\ref{musc}) are independent
of $N_c$. Natural candidates are, for instance, quark loops,
like in the NJL model. We therefore expect that the
quark-meson loop model of Ref.\ \cite{Schaefer:2007pw} shows
the correct large-$N_c$ scaling of $T_c$.

Our second suggestion for improving the
linear $\sigma$-model is to incorporate the coupling to
the Polyakov loop, defined as 
$$
l(x)=N_{c}^{-1}\mathrm{Tr}\left[
\mathcal{P}~\text{exp}\left(  \imath g_{QCD}\int_{0}^{1/T}A_{0}(\tau
,x)d\tau\right)  \right]\; ,
$$ 
where the trace runs over all color degrees of
freedom, $\mathcal{P}$ stands for path ordering, and $A_{0}(\tau,x)$ is
the zero component of the gluon field $A_{\mu}$ \cite{Polyakov:1978vu}. In
pure gauge theory the expectation of the Polyakov loop $l(T)=\left\langle
l(x)\right\rangle $ is an order parameter for the deconfinement phase
transition: $l=0$ in the confined phase and $l=1$ in the deconfined
phase, see the review in Ref.\ \cite{Fukushima:2010bq} and refs.\ therein.

Following Ref.\ \cite{Dumitru:2000in} (for a similar approach see also Ref.\ \cite{Mocsy:2003qw}) 
we couple the $\sigma$-model to the Polyakov loop
\[
\mathcal{L}_{\sigma\text{-Pol}}(N_{c})=\mathcal{L}_{\sigma}(N_{c}%
)+\frac{\alpha N_{c}}{4\pi}|\partial_{\mu}l|^{2}T^{2}-\mathcal{V}%
(l)-\frac{h^{2}}{2}\,\Phi^{2}|l|^{2}T^{2}\text{ .}%
\]
where $\mathcal{L}_{\sigma}(N_{c})$ is taken from Eq.\ (\ref{ls}) and the
Polyakov loop is coupled to the meson fields. Moreover, a kinetic term and a
potential $\mathcal{V}(l)$ for the Polyakov field $l$ have been introduced.
(Since we are only interested in the large-$N_{c}$ behavior, the precise form
of $\mathcal{V}(l)$ is irrelevant in the following. Terms of the kind $\sim l T \Phi^2$
could also be included \cite{Mocsy:2003qw} but would not affect the overall $N_c$-scaling, although
they might change the order of the phase transition.) Applying the CJT formalism in
double-bubble
approximation the gap equation for the condensate $\varphi(T)$ now reads
\[
0= \varphi(T)^{2}-\frac{N_c}{3 \lambda}\, \left( \mu^{2}
- h^2 |l|^{2}T^{2} \right)
+3 \int\left(  G_{\sigma}+G_{\pi}\right) \;,
\]
from which the following expression for the
critical temperature $T_{c}$ is derived:
\begin{equation}
T_{c}=\frac{\mu}{\sqrt{h^{2}|l(T_{c})|^{2}+\frac{6\lambda}{N_{c}}}}\text{ .}%
\end{equation}
Assuming that $l(T_c)$ is a constant independent
of $N_c$, we again obtain $T_c \propto N_c^0$ in 
the limit $N_{c}\rightarrow\infty$.
Detailed numerical results represent a task for the
future. They depend on the form of the Polyakov-loop potential and on other
parameters of the model. However, the important point here is that it
is natural to recover the desired large-$N_{c}$ limit when the hadronic model
is coupled to the Polyakov loop. The reason for this is
that the chiral phase transition is triggered by the Polyakov loop
\cite{Meisinger:1995ih}.


In conclusion, we have shown that models of NJL-type and linear $\sigma$-type predict a
different behavior of $T_{c}$ as function of $N_{c}$. In the quark-based NJL
model $T_{c}$ is independent of $N_{c}$: this result agrees with the
expectation that $T_{c}$ scales as $\Lambda_{QCD}\propto N_{c}^{0},$ just as
the deconfinement phase transition. On the contrary, in linear $\sigma$-type models
a scaling $T_{c}\propto\sqrt{N_{c}}$ is obtained. The different scaling originates
from the particular mechanism which restores chiral symmetry 
in the two models. In the NJL model,
quark loops are responsible for chiral symmetry restoration,
which survive in the large-$N_{c}$ limit, while in the linear
$\sigma$-model, meson loops induce the chiral phase transition, which disappear in the
large-$N_{c}$ limit.

This mismatch is, at first sight, even more striking because the
linear $\sigma$-model can be derived from the NJL model through a hadronization procedure.
However, since one should perform this hadronization 
at each $T,$ the parameters of the $\sigma$-model are actually
functions of $T.$ Indeed, when making (at least) one parameter of the
$\sigma$-model $T$-dependent, the expected large-$N_{c}$ limit can be easily
recovered. This result, although interesting, is based on an {\it ad hoc\/}
modification of the $\sigma$-model: for this reason we have also studied a
different approach, in which --inspired by Ref.\ \cite{Dumitru:2000in}-- we
have coupled the linear $\sigma$-model to the Polyakov loop. In this way the chiral
phase transition is induced by the deconfinement phase transition and, as a
consequence, $T_{c}$ scales as $N_{c}^{0}.$

We therefore conclude that the study of the chiral phase transition within
purely hadronic models is possible, although care is needed in order to be in
agreement with the large-$N_{c}$ limit. Detailed numerical studies with more
realistic models including (pseudo)scalar and (axial-)vector mesons will be
presented in the future.

\bigskip

\textbf{Acknowledgment: }The authors thank E.\ Seel, M.\ Grahl and  R. Pisarski for useful
discussions. A.H.\ thanks H-QM and HGS-HIRe for financial support.

\end{document}